\documentclass[prl,aps,twocolumn,showpacs,superscriptaddress]{revtex4}
\usepackage{bm}
\usepackage{amsmath,amssymb}
\usepackage{graphicx}

\begin{document}

\title{A high-accuracy optical clock via three-level coherence in
neutral bosonic $^{88}$Sr}
\author{Robin Santra}
\affiliation{JILA, National Institute of Standards and Technology and
University of Colorado\\
Department of Physics, University of Colorado, Boulder, Colorado 80309-0440, USA}
\affiliation{ITAMP, Harvard-Smithsonian Center for Astrophysics,
Cambridge, MA 02138, USA}
\author{Ennio Arimondo}
\affiliation{JILA, National Institute of Standards and Technology and
University of Colorado\\
Department of Physics, University of Colorado, Boulder, Colorado 80309-0440, USA}
\affiliation{INFM UdR di Pisa, Dipartimento di Fisica E. Fermi,
Universit\`a di Pisa, Via Buonarroti 2, I-56127 Pisa, Italy}
\author{Tetsuya Ido}
\affiliation{JILA, National Institute of Standards and Technology and
University of Colorado\\
Department of Physics, University of Colorado, Boulder, Colorado 80309-0440, USA}
\author{Chris H. Greene}
\affiliation{JILA, National Institute of Standards and Technology and
University of Colorado\\
Department of Physics, University of Colorado, Boulder, Colorado 80309-0440, USA}
\author{Jun Ye}
\affiliation{JILA, National Institute of Standards and Technology and
University of Colorado\\
Department of Physics, University of Colorado, Boulder, Colorado 80309-0440, USA}

\date{\today}
\begin{abstract}
An optical atomic clock scheme is developed that utilizes two
lasers to establish coherent coupling between the $5s^2 \;
\phantom{}^1S_0$ ground state of $^{88}$Sr and the first excited
state, $5s5p \; \phantom{}^3P_0$. The coupling is mediated by the
broad $5s5p \; \phantom{}^1P_1$ state, exploiting the phenomenon
of electromagnetically induced transparency. The effective
linewidth of the clock transition can be chosen at will by
adjusting the laser intensity. By trapping the $^{88}$Sr atoms in
an optical lattice, long interaction times with the two lasers are
ensured; Doppler and recoil effects are eliminated. Based on a
careful analysis of systematic errors, a clock accuracy of better
than $2\times 10^{-17}$ is expected.
\end{abstract}
\maketitle

Armed with superior resonance quality factors, optical atomic
clocks based on single trapped ions or a collection of laser
cooled neutral atoms are expected to outperform microwave-based
atomic clocks in the near future~\cite{Didd01,Oate99,Wilp02}.
While a large number of quantum absorbers in a neutral atom
system provide an advantage in the enhanced short-term frequency
stability, atomic motion during the probe phase seriously limits
the attainable accuracy. A competitive proposal is to localize
neutral cold atoms spatially in a Lamb-Dicke regime while the
trapping potential is designed such that its presence does not
influence the clock transition frequency~\cite{IdKa03}. This
scheme can be potentially realized using a far-off-resonance
dipole trap operating at a wavelength where the ground and the
excited state of the clock transition experience exactly the same
AC Stark shifts. In particular, if the atoms are confined near
the ground vibrational levels in an optical lattice, frequency
shifts associated with effects such as Doppler, collision, and
recoil will all be reduced to negligible levels.

To realize such a scheme, it is of essential importance that the
polarizabilities of the two clock states are matched to a high
degree of accuracy at the ``magic'' wavelength~\cite{KaTa03}.
States with scalar polarizabilities are preferred, to avoid the
problem of the complex and sometimes uncontrolled light
polarization inside an optical lattice. The fermionic strontium
isotope, $^{87}$Sr, offers a nearly satisfactory
solution~\cite{KaTa03,TaKa03}. The nuclear magnetic dipole moment
of $^{87}$Sr makes the $5s^2 \; \phantom{}^1S_0$ -- $5s5p \;
\phantom{}^3P_0$ transition weakly dipole allowed, with a
predicted linewidth of about 1~mHz~\cite{KaTa03,SaCh04,PoDe04}.
The two clock states have very small electronic angular momenta
(due to hyperfine mixing) and their polarizabilities are nearly
scalar. However, the hyperfine structure can cause a clock
frequency shift (possibly as large as 10 Hz) by fluctuations of
the lattice light polarization. The large nuclear spin ($I =
9/2$) also brings complexity in state preparation and field
control.

In this Letter, we propose a scheme for an optical clock that is
based on states of true scalar nature, namely $5s^2 \;
\phantom{}^1S_0$ -- $5s5p \; \phantom{}^3P_0$ of $^{88}$Sr (with
$I=0$). A direct transition between these two states is of course
completely forbidden, and the clock scheme is based instead on
three-level quantum coherence established by two probe laser
frequencies. We note that recently a scheme has emerged based on
four levels and three probe lasers~\cite{HoCr04}. An estimate of
the expected accuracy for this scheme is not yet available. A
scheme similar to what is discussed here was investigated for
trapped ions~\cite{siemers}.

\begin{figure}[h]
\includegraphics*[width=5.2cm,origin=c,angle=0]{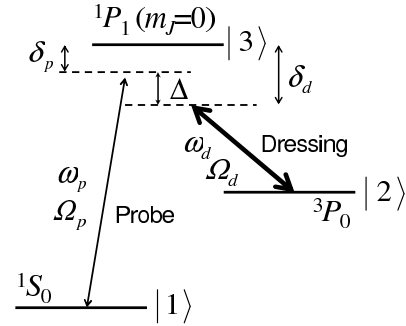}
\caption[]{Three-level coherence in neutral $^{88}$Sr. $5s^2 \;
\phantom{}^1S_0$ and $5s5p \; \phantom{}^3P_0$ are the two clock
levels. By applying the probe and the dressing laser as shown, a
coherent coupling is established between the clock levels,
mediated by the $5s5p \; \phantom{}^1P_1$ level.} \label{fig1}
\end{figure}

Our proposal rests on the three-level lambda-type scheme shown in
Fig.~\ref{fig1}. Particularly promising in this context is
strontium~\cite{IdKa03,TaKa03,KaTa03,XuLo03}, which we consider
here in detail. A similar scheme is conceivable for bosonic
isotopes of other alkaline-earth species (including ytterbium).
Since the ground state ($5s^2 \; \phantom{}^1S_0$)---subsequently
denoted by the state vector $\left|1\right\rangle$---and the
first excited state ($5s5p \;
\phantom{}^3P_0$)---$\left|2\right\rangle$---of $^{88}$Sr are
characterized by a total angular momentum of $J=0$, electric and
magnetic one-photon transition matrix elements between these two
states vanish to any multipole order. $\left|2\right\rangle$
decays radiatively via E1M1 two-photon emission after a lifetime
of a few thousand years~\cite{SaCh04}. For practical purposes, it
is therefore legitimate to set the decay width of
$\left|2\right\rangle$ equal to zero. The $5s5p \;
\phantom{}^1P_1$ state, which is referred to as
$\left|3\right\rangle$ in our scheme, can be reached from
$\left|1\right\rangle$ via E1 one-photon absorption using a {\em
probe} laser at a wavelength of $461$ nm (frequency $\omega_p$).
One M1 photon from a {\em dressing} laser at $1354$ nm (frequency
$\omega_d$) is needed to resonantly couple $\left|2\right\rangle$
and $\left|3\right\rangle$. The latter decays primarily to
$\left|1\right\rangle$. The decay width is $\gamma=2\pi\times 32$
MHz.

Assuming linear polarization of both the probe and the dressing
lasers, and assuming that the electric component of the probe
laser (amplitude $\mathcal{E}_0$) and the magnetic component of
the dressing laser (amplitude $\mathcal{B}_0$) are aligned along
the same axis, level $3$ is characterized by $m=0$ for our
purposes. Let $\Omega_p = \left\langle 3 \right|D_0\left| 1
\right\rangle \mathcal{E}_0/\hbar$ denote the Rabi frequency
associated with electric-dipole coupling between levels $1$ and
$3$. Similarly, $\Omega_d = \left\langle 3 \right|M_0\left| 2
\right\rangle \mathcal{B}_0/\hbar$ is the Rabi frequency
referring to the dressing process. $\Omega_p$ and $\Omega_d$ may
be taken to be real.

We first analyze the dressed-atom problem within the subspace spanned by
$\left|2\right\rangle$ and $\left|3\right\rangle$ in the absence of the
probe laser. The equations of motion in the interaction picture read
\begin{eqnarray}
\label{eq1}
\mathrm{i} \dot{c}_2(t) & = & \frac{\Omega_d}{2} \mathrm{e}^{\mathrm{i} \delta_d t}
c_3(t) \; , \\
\label{eq2}
\mathrm{i} \dot{c}_3(t) & = & \frac{\Omega_d}{2} \mathrm{e}^{-\mathrm{i} \delta_d t}
c_2(t) -\mathrm{i} \frac{\gamma}{2} c_3(t) \; ,
\end{eqnarray}
where $c_2$($t$) and $c_3$($t$) represent probability amplitudes
and the rotating-wave approximation has been applied. The
dressing laser detuning is $\delta_d = \omega_d - (\omega_3 -
\omega_2)$ [$\omega_i$ stands for the unperturbed frequency of
$\left|i\right\rangle$, $i=1,2,3$].

Due to the action of $\Omega_d$, some $\left|3\right\rangle$ character is admixed to
$\left|2\right\rangle$ (and vice versa), so that a long-lived dressed state,
$\left|\tilde{2}\right\rangle$, emerges whose decay rate can be adjusted through the
dressing laser intensity. It is easily seen that to leading order with respect to
$\Omega_d/(\delta_d + \mathrm{i} \gamma/2)$
\begin{equation}
\label{eq3}
\left|\tilde{2}\right\rangle =
\left\{
\left|2\right\rangle \mathrm{e}^{-\mathrm{i} \omega_2 t}
+ b \left|3\right\rangle \mathrm{e}^{-\mathrm{i} (\omega_3+\delta_d) t}
\right\} \mathrm{e}^{-\mathrm{i} \beta t} \; .
\end{equation}
Here,
\begin{equation}
\label{eq4}
b = \frac{1}{2}\frac{\Omega_d}{\delta_d + \mathrm{i} \gamma/2} \; \; , \; \;
\beta = \frac{1}{2} \Omega_d b \; .
\end{equation}
It follows from the equation for $\beta$ that for small detuning
the decay rate of $\left|\tilde{2}\right\rangle$ is
$\tilde{\gamma} = \Omega_d^2/\gamma$. The dressed state deriving
primarily from $\left|3\right\rangle$ decays with a rate that is
virtually identical to $\gamma$.

Using the atomic structure code described in Ref.~\cite{SaCh04},
which accurately treats electron correlation and spin-orbit
effects in the valence shell, we obtain for the magnetic coupling
matrix element $\left|\left\langle 3 \right|M_0\left| 2
\right\rangle\right| = 0.022$ $\mu_{\mathrm{B}}$. Hence, if an
effective width $\tilde{\gamma}$ of, say, 1 mHz is desired
(similar to the $^{87}$Sr case), a dressing laser intensity of 3.9
mW/cm$^2$ must be chosen and is used for numerical results
presented in this paper. (By decreasing the intensity, the
effective linewidth can be made even narrower.) Under these
circumstances, the parameter $b$ in Eqs.~(\ref{eq3}) and
(\ref{eq4}), which describes the degree of admixture of
$\left|3\right\rangle$ character in the long-lived dressed state,
has a magnitude of less than $10^{-5}$. In other words, the
dressed states in this scheme are hardly distinguishable from
bare atomic states. We will exploit this fact later when
estimating level shifts caused by sources not included in the
simple three-level picture of Fig.~\ref{fig1}.

Since there is no direct coupling between $\left|1\right\rangle$
and $\left|2\right\rangle$, it is apparent from Eq.~(\ref{eq3})
that the transition associated with the long-lived dressed state
can only be probed from the ground state by tuning $\omega_p$ in
the vicinity of $\omega_3 + \delta_d - \omega_1$, i.e. in the
vicinity of the broad absorption profile of level 3. Hence,
$\delta_p = \omega_p - (\omega_3 - \omega_1)$ $\sim$ $\delta_d$.
The long-lived and the short-lived states are excited in a
coherent fashion by the probe laser. They interfere and give rise
to a narrow dip structure in the absorption profile---an effect
known as {\em electromagnetically induced transparency}
(EIT)~\cite{Harr97}. This dip is more than $10^{10}$ times
narrower ($\tilde{\gamma} = 2\pi \times 1$ mHz) than the bare
absorption profile of level 3. The broad background is therefore
essentially constant across the narrow structure.

This quantum interference may be treated, to a first
approximation, within a wave function-based approach, but it fails
to take into consideration the repopulation of
$\left|1\right\rangle$ via spontaneous emission from
$\left|3\right\rangle$. More importantly, the loss of coherence
between levels $1$ and $2$ cannot be described. Decoherence is
caused by lattice-induced spontaneous emission, atomic tunneling
between adjacent sites in the optical lattice and ensuing
collisions between the atoms. Another factor is the loss of phase
coherence between the probe and the dressing laser. Both aspects
must be included in a realistic description. Utilizing standard
density-matrix theory~\cite{BrHa75}, the absorption rate from the
ground state is found to be
\begin{widetext}
\begin{equation}
\label{eq5}
W_{\mathrm{abs}} = \frac{\Omega_p^2}{\gamma}
\frac{\Delta^2+\gamma_c \{\gamma_c+\tilde{\gamma}/2\}}{[2 \Delta(\Delta+\delta_d)/\gamma
- \{\gamma_c + \tilde{\gamma}/2\}]^2 +
[\Delta+2\gamma_c(\Delta+\delta_d)/\gamma]^2} \; ,
\end{equation}
\end{widetext}
where $\Delta=(\omega_p - \omega_d) - (\omega_2 - \omega_1)$, and
$\gamma_c$ stands for the loss rate of coherence between
$\left|1\right\rangle$ and $\left|2\right\rangle$.

\begin{figure}[h]
\includegraphics[width=8.0cm,origin=c,angle=0]{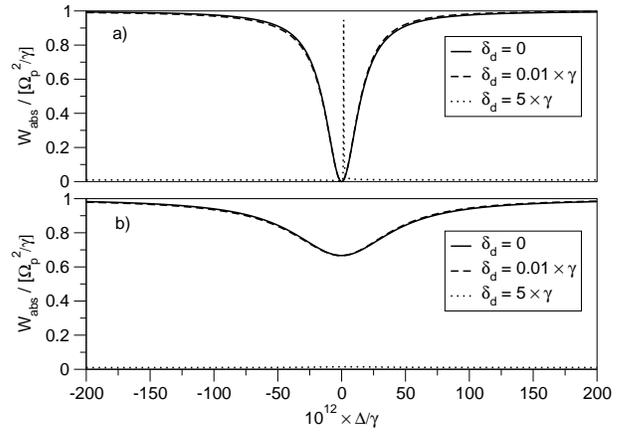}
\caption[]{Absorption rate from the ground state of $^{88}$Sr in
the presence of the two laser frequencies indicated in
Fig.~\ref{fig1}. The laser-induced width, $\tilde{\gamma}$, is
$2\pi\times 1$ mHz. a) Loss of phase coherence between levels 1
and 2 is neglected, $\gamma_c=0$. Note that the curves for
$\delta_d=0$ and $\delta_d=0.01\times\gamma=2\pi\times 320$ kHz
are virtually indistinguishable. b) Macroscopic decoherence and
laser-induced decay are comparable, $\gamma_c=\tilde{\gamma}$.}
\label{fig2}
\end{figure}

If the dressing laser is off, or if $\gamma_c \gg
\tilde{\gamma}$, the simple Lorentzian profile connected with the
transition from $\left|1\right\rangle$ to $\left|3\right\rangle$
is recovered. Otherwise, if we disregard $\gamma_c$ for a moment,
the absorption rate vanishes at $\Delta=0$. Note that the
occurrence of a zero in this absorption profile is clear from the
connection of this problem to the Fano~\cite{Fano61} lineshape
for one bound state (level 2 upshifted by the energy of one
dressing laser photon) embedded in one ``continuum'' (the 32 MHz
wide $\left|3\right\rangle$ state). A deep, narrow dip occurs in
the absorption profile ~\cite{note}, and the width of the EIT dip
for $\gamma_c=0$ is
\begin{equation}
\label{eq6}
\Delta_{\mathrm{FWHM}} = \tilde{\gamma}
\left\{
1 - 8\frac{\delta_d^4}{\gamma^4} + O\left(\frac{\delta_d^6}{\gamma^6}\right)
\right\} \; .
\end{equation}
This demonstrates that not only the position of the EIT dip, but
also its width is very insensitive with respect to fluctuations
of $\delta_d$. A variation of the dressing laser frequency by as
much as 100 kHz induces a relative change of the EIT width of
less than $10^{-7}$. The effect is illustrated in
Fig.~\ref{fig2}(a) ($\gamma_c=0$). Hence, by measuring
$W_{\mathrm{abs}}$ as a function of $\omega_p - \omega_d$, the
clock frequency $\omega_2 - \omega_1$ can be determined with high
accuracy by tracking the position of the dip. Neither $\omega_p$
nor $\omega_d$ must be particularly stable, only the difference
needs to be stabilized.

Also shown in Fig.~\ref{fig2}(a) is the absorption profile near
$\Delta=0$ for relatively large detuning
($\delta_d=5\times\gamma$). Overall, the signal amplitude is
drastically reduced in this case. There is a narrow absorption
peak in the immediate vicinity of $\Delta=0$, but, as can be seen
in Fig.~\ref{fig2}(b), this peak all but disappears as soon as
decoherence becomes appreciable. (We have chosen
$\gamma_c=\tilde{\gamma}$ in Fig.~\ref{fig2}(b).) On the other
hand, for small detuning the EIT dip is still clearly visible,
even though it is not as deep as in the absence of decoherence
and it has become broader. The position of the EIT minimum and
the insensitivity of the EIT width with respect to laser
frequency ($\delta_d$ or $\delta_p$) fluctuations of 100 kHz or so
are essentially unaffected.

The discussion has so far covered only resonant couplings within
the span of $\left|1\right\rangle$, $\left|2\right\rangle$, and
$\left|3\right\rangle$. In order to evaluate the potential
accuracy of the proposed clock scheme, we need to consider the AC
Stark shifts due to nonresonant electric dipole coupling of
$\left|1\right\rangle$, $\left|2\right\rangle$, and
$\left|3\right\rangle$ to states inside and outside the
three-level subspace. Our calculation of these shifts employs
accurate experimental and theoretical data on excitation energies
and oscillator strengths of atomic
Sr~\cite{IdKa03,WeGr92,UeAs82}. For a probe laser intensity of
10~$\mu$W/cm$^2$, levels 1 and 2 are AC Stark-shifted by less
than 1~mHz. The dressing laser, operating at an intensity of
3.9~mW/cm$^2$, causes an AC Stark shift of levels 1 and 2 of
-41~mHz and -20~mHz, respectively. Thus, the clock frequency is
systematically shifted by about +19~mHz. Since the intensity of
the dressing laser can be stabilized to better than 1~\%, we
conclude that the AC Stark effect due to the probe and dressing
lasers can be experimentally characterized at the sub-mHz level.

To allow for a sufficiently long interrogation time and eliminate
systematic frequency shifts associated with atomic motion, the
atoms will be confined in a Lamb-Dicke regime in an optical
lattice at the magic wavelength for the clock transition, as
shown in Fig. 3(a). The lattice trapping field causes, to leading
order, no net frequency shift of the clock transition as the
level shifts of 1 and 2 (150 kHz well depth in Fig. 3(a), under a
lattice laser intensity of 10 kW/cm$^2$) are exactly matched. (The
studies in Ref.~\cite{KaTa03} suggest that higher-order AC Stark
effects are negligible at this lattice laser intensity.) Level 3
in our scheme, however, is shifted relative to levels 1 and 2,
and this shift depends on the local intensity an atom experiences
in the lattice. We calculate a relative shift of -250~kHz
(maximum in magnitude), at a lattice laser intensity of
10~kW/cm$^2$ and a magic wavelength of 813.5~nm~\cite{TaKa03}.
Figure 3(a) shows the lattice-induced shift of level 3 in
comparison to $\gamma$. This shift is similar---in effect and in
magnitude---to the drift of the probe and dressing laser
frequencies. It has, therefore, an equally negligible effect on
the properties of the EIT dip.

\begin{figure}[h]
\includegraphics[width=8cm,origin=c,angle=0]{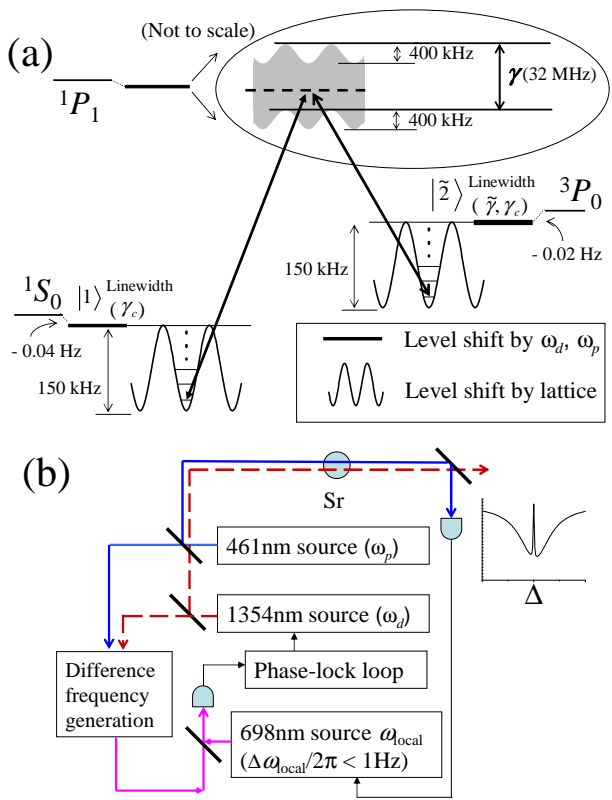}
\caption[]{(a) Lattice-confined atoms for use in the EIT clock
scheme. The AC Stark shift due to the dressing laser on the clock
transition is shown, as are the level shifts associated with the
lattice trapping field. The expanded view of the $^1P_1$ state
compares the lattice-induced level shift versus the linewidth
$\gamma$. Atoms are cooled to the Lamb-Dicke regime, predominantly
occupying the ground vibrational level in lattice. (b)
Experimental implementation of the EIT-based optical clock. A
pre-stabilized narrow-linewidth local oscillator at 698 nm
establishes a tight frequency track between the probe and
dressing lasers.  This enables a probe of the narrow EIT
resonance, which in turn provides a long-term clock reference for
the 698 nm local oscillator. } \label{fig3}
\end{figure}

To make the final estimate of the clock accuracy, the level
shifts shown in Fig. 3(a) contain the AC Stark shifts due to the
simultaneous presence of $\omega_p$ and $\omega_d$ laser fields
as discussed above. The lattice field induces a spontaneous
emission limited linewidth of the clock transition at $\sim$15
mHz. The detuning and decoherence-(laser frequency variations,
atomic motion inside lattice, spontaneous emission, etc)-related
changes in the EIT linewidth may lead to a small shift in the
experimental search of the linecenter, but the shift magnitude is
estimated to be under $10^{-18}$. The blackbody radiation induced
clock frequency shift is estimated to be -0.54 Hz at 300 K, with
a temperature coefficient of -7.5 mHz/K. Finally, the strict
scalar nature of the clock transition reduces the light
polarization and stray magnetic field-induced frequency shift
below $10^{-18}$. In view of the available control accuracy of
the laboratory temperature and the dressing laser intensity, the
overall accuracy limit is estimated to be smaller than 8 mHz, or
$2\times 10^{-17}$.

Experimental implementation of this EIT-based clock scheme is
shown in Fig. 3(b). A narrow linewidth ($<$ 1 Hz) laser at 698 nm
($\omega_{local}$) will be developed by prestabilizing
$\omega_{local}$ to a stable, passive optical cavity. The
difference frequency, $\omega_p - \omega_d$, between the probe
laser and the dressing laser, generated by a nonlinear optical
crystal, will be phase locked to the 698 nm local oscillator. One
of the two lasers, such as $\omega_d$, as shown in Fig. 3(b), can
be used as the slave to accomplish the optical phase lock loop.
Thus, $\omega_p - \omega_d$ = $\omega_{local}$. This is
prestabilization, implemented before $\omega_p$ and $\omega_d$
interact with the atoms. $\omega_p$ and $\omega_d$ may fluctuate
on the order of 100 kHz, but their difference fluctuates less
than 1 Hz.

The frequency of $\omega_{local}$ can be precisely scanned with
reference to a stable cavity mode, which of course has an
inevitable slow frequency drift due to the material nature of the
reference. However, precise scanning of $\omega_p - \omega_d$
through the EIT resonance enables one to activate a slow feedback
loop to stabilize $\omega_p - \omega_d$ to the value of
$\omega_2 - \omega_1$, i.e., the quantum resonance corrects the
long-term drift of the cavity resonance according to the clock
transition frequency. The feedback is applied directly on the 698
nm local oscillator ($\omega_{local}$) to ensure $\Delta$ = 0.

The tunability of the linewidth, the exquisite insensitivity with
respect to light polarization in the optical lattice, the
straightforward state control, and the small, controllable AC
Stark shifts promise that our EIT-based clock scheme using bosonic
$^{88}$Sr will represent a practical and robust approach for
optical atomic clocks with a competitive performance in both
stability and accuracy.

\acknowledgments We thank M. Boyd, A. Ludlow, T. Loftus, J. Hall,
S. Yelin, and M. Lukin for stimulating discussions. Funding is
provided by NSF, NIST, ONR, and NASA. J. Ye's email
Ye@jila.colorado.edu.


\begin{thebibliography}{99}
\bibitem{note} The absence of AC Stark shifts associated with the
dressing and probe lasers acting within the three level system was
pointed out in \cite{siemers} and references therein.
\end{thebibliography}
\end{document}